# A combined segmented anode gas ionization chamber and time-of-flight detector for heavy ion elastic recoil detection analysis

*Petter Ström*[1]

*Per Petersson*[1]

*Marek Rubel*[1]

*Göran Possnert*[2]

[1]Department of Fusion Plasma Physics, School of Electrical Engineering, KTH Royal Institute of Technology, SE-10044 Stockholm, Sweden

[2]Department of Physics and Astronomy, Ion Physics, Uppsala University, P.O. Box 516, SE-75120 Uppsala, Sweden

## Abstract

A dedicated detector system for heavy ion elastic recoil detection analysis at the Tandem Laboratory of Uppsala University is presented. Benefits of combining a time-of-flight measurement with a segmented anode gas ionization chamber are demonstrated. The capability of ion species identification is improved with the present system, compared to that obtained when using a single solid state silicon detector for the full ion energy signal. The system enables separation of light elements, up to Neon, based on atomic number while signals from heavy elements such as molybdenum and tungsten are separated based on mass, to a sample depth on the order of one micrometer. The performance of the system is discussed and a selection of material analysis applications is given. Plasma-facing materials from fusion experiments, in particular metal mirrors, are used as a main example for the discussion. Marker experiments using nitrogen-15 or oxygen-18 are specific cases for which the described improved species separation and sensitivity are required. Resilience to radiation damage and significantly improved energy resolution for heavy elements at low energies are additional benefits of the gas ionization chamber over a solid state detector based system.

## 1. Introduction

Time-of-flight heavy ion elastic recoil detection analysis (ToF-HIERDA) [1] is widely used to perform depth-profile measurements of elemental composition in the surface layer of material samples and for studying thin films [2]. With a depth range on the order of one micrometer and depth resolution on the order of 10 nm in the absence of surface roughness, the technique can be used to examine for example dopant migration in lattices under annealing [3], the chemical composition of coatings [4] and materials used in electronics [5]. ToF-HIERDA has also been applied to the study of material modification on plasma-facing components in fusion experiments [6-8]. For this application, it is of importance to obtain a clear separation between signals from different relevant ion species for light and heavy elements simultaneously. In particular the signals from W and Mo should be separated from that of scattered heavy ions originating from the primary beam, typically $^{127}I^{8+}$ [9], while at the same time all isotopes in the mass range from H to Ne must be correctly identified. Such identification is needed when studying materials relevant for the International Thermonuclear Experimental Reactor (ITER) [10], since W will be exclusively used as plasma-facing material in the divertor region of that machine [11] and Be will be present in the main chamber wall [12]. Mo is used as the main material for diagnostic mirrors which are being extensively studied with ToF-HIERDA for the First Mirror Test program at the Joint European Torus (JET) since 2006 [13,14]. Furthermore, measurements on samples of fusion reactor wall materials often follow tracer experiments [15], where a range of light elements are to be quantified individually [16].

Systems with gas ionization chambers (GIC) have previously been constructed to provide optimal ion species separation in specific mass-ranges, either measuring the full energy of incoming ions and combining it with a time-of-flight (ToF) measurement [17,18], or differentiating those ions based on stopping power in the detection gas (so-called ΔE/E). The latter is exemplified in ref. [19] for an application in accelerator mass spectrometry. We have made use of both these methods in a ToF-GIC setup in order to provide a versatile system with a good overall energy resolution that can be tuned for enhanced ion species separation in a specified range of mass, nuclear charge number and energy.

## 2. Detector setup

A top view sketch of the detection system as installed at the Tandem accelerator laboratory of Uppsala University is shown in Fig. 1. Samples are placed in a chamber at the right side of the depicted region. They can be rotated freely around the central axis of a manipulator. For standard measurements the angle between the sample surface and the analyzing beam is 23±1°. The detection system is fixed at 45±1° with respect to the forward beam direction. In order to provide the ability to study larger samples without having to cut them into pieces, the manipulator can be moved 20 cm in the "up-down" direction, perpendicular to the beam, and approximately 3 cm in each of the remaining two directions (parallel and perpendicular to the beam). Assuming that the angle between the beam and the sample surface is 22.5°, the up-down and sideways movement provides a rectangle on the surface of 20x7.8 $cm^2$ which can be analyzed. Although the most common sample size is on the order of 1x1 $cm^2$, samples with surface areas up to approximately 20x15 $cm^2$ and 2 cm thickness can be mounted in the sample chamber. Two ToF detectors are present, labelled ToF 1 and ToF 2 in the figure. The GIC is shown farthest to the left. The figure is drawn to scale, apart from the sample size which is exaggerated. Both ToF detectors can be retracted from the path of incoming ions using manipulator handles on top of the respective vacuum chambers.

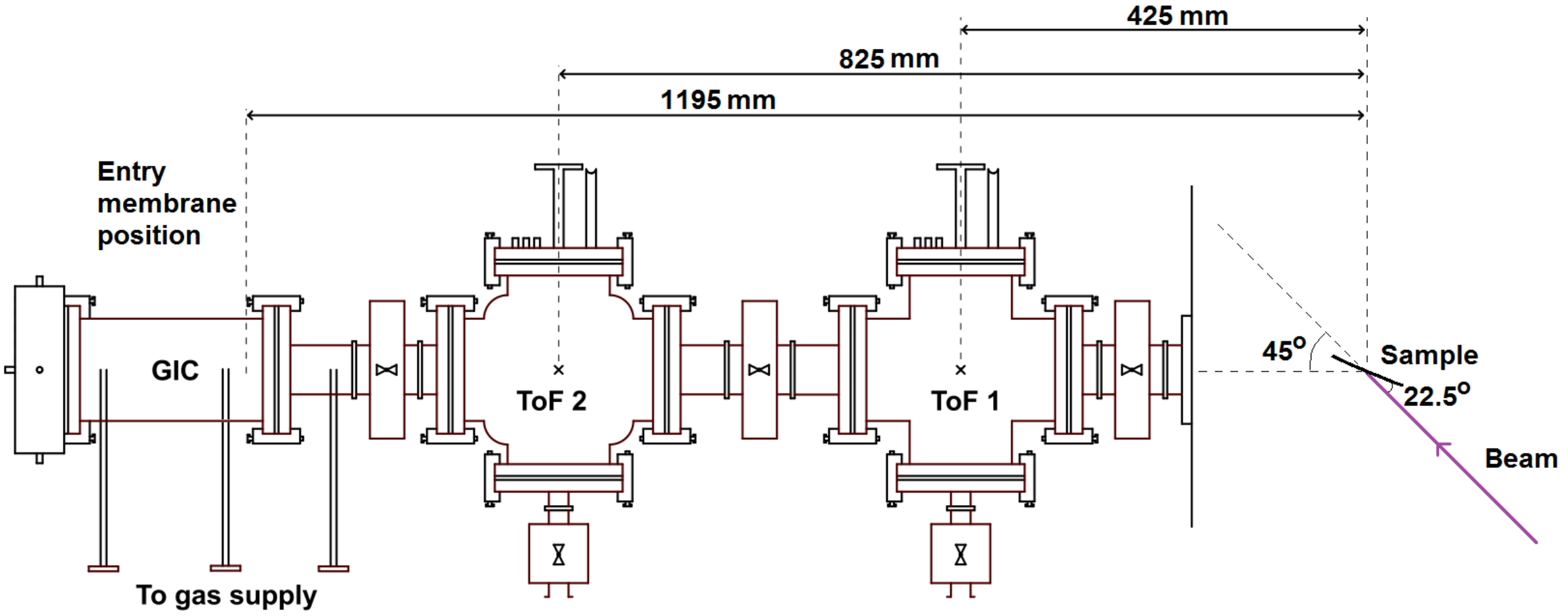

**Figure 1:** Sketch of the detection system, seen from above, with distances from the ideal sample position to the time-of-flight detectors and the entry membrane of the GIC. The valves depicted below the ToF detectors are used to vent and pump the individual parts, without breaking vacuum in the entire system, if component replacement is necessary. The ToF detectors have been rotated by 90 degrees in the figure, and the manipulator handles are in reality pointing upwards.

*2.1 Time-of-flight*

The ToF measurement is based on start and stop signals recorded with two identical detectors following the design described in ref. [20]. Ions to be detected pass through a circular carbon foil of 5 μg/cm$^2$ and 10 mm radius. Electrons that are ejected as a result of interaction in the foil are accelerated towards a pair of microchannel plates (MCP) in chevron stack configuration. Pairwise resistance-matched MCPs from Hamamatsu Photonics, model F1094-011 [21], were selected for this purpose. The measurements presented here, however, use older MCPs from Philips Photonics (presently Photonis) with similar specifications. The electrical connection details for the ToF detectors are shown in Fig. 2. A high voltage, typically 4200 V, is divided over a series of resistors and applied to consecutive grids forcing the electrons towards the MCPs. The final stages of the resistive divider supply the voltages across the MCPs themselves, which assuming 30 MΩ plate resistance (average from ref. [21]), is 730 V per MCP. Finally, the output signal is taken from an aluminum plate below the second MCP with a 50 Ω resistor to ground for impedance matching and amplified with a wideband preamplifier, model 6954 from Phillips Scientific. A constant fraction discriminator, model 6915 from Phillips Scientific, is used to remove the dependency of the trigger time for subsequent electronics on variations in the pulse height of the preamplifier output, providing a time resolution of approximately 300 ps.

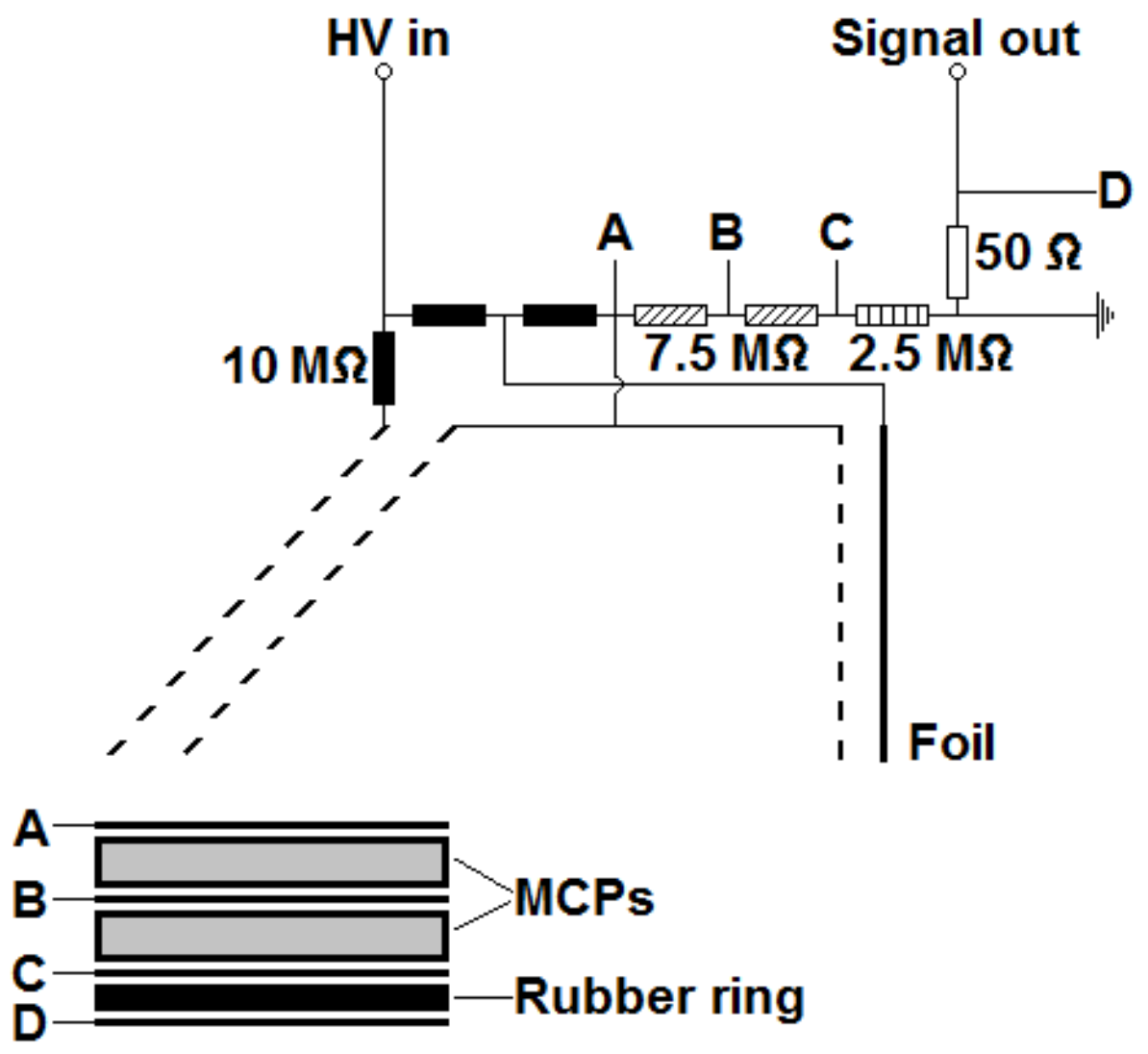

**Figure 2:** Electronic configuration of the ToF detectors. The letters A-D correspond to points that are electrically connected (A to A, B to B etc.), and the dashed lines represent the grids used to guide electrons to the MCPs. Resistors are pattern-coded to indicate their resistance (black fill: 10 MΩ, diagonally striped: 7.5 MΩ, vertically striped: 2.5 MΩ, white fill: 50 Ω).

From the point of measurement on the sample surface, the distance to the first ToF-foil is 425 mm, to the second ToF-foil 825 mm and to the entry window of the GIC 1195 mm. Consequently, the flight distance for the ToF measurement is 400 mm and the solid angle coverage of the GIC window is 0.057 msr. In order to avoid counting ions that only hit the first ToF foil, but miss the second, the output from the second ToF detector (the one farthest from the sample) is used as starting signal for an Ortec 567 time-to-amplitude converter. The output from the first ToF detector is delayed using 58.0 m of RG58 cable and used as stopping signal. With a signal propagation speed of 0.659c in the cable, this sets an upper boundary to the flight time of ions to be detected at approximately 294 ns. For W, the corresponding lower energy threshold for detection is 1.76 MeV, whereas for hydrogen it is 9.65 keV, the latter of which is below the energy resolution of the GIC. As such, the length of the delay cable is sufficient to not cause any loss of low energy hydrogen counts.

### *2.2 Gas ionization chamber*

The GIC constitutes the last detection stage. Its outer shell is a cylindrical pipe of stainless steel with 100 mm inner diameter and a length of 260 mm. Isobutane ($C_4H_{10}$) is used as detection gas inside the cylinder with a silicon nitride membrane from Silson Ltd. separating it from the vacuum that is maintained in the rest of system. The area of the membrane is 10x10 mm$^2$, with two silicon support ribs of 0.5 mm width and 200 µm thickness in each direction to help sustain the pressure gradient. The active membrane area is thus 9x9 mm$^2$. A membrane thickness of 75 nm and gas pressures below 50 mbar were used for the measurements presented here, but the membrane can be replaced if higher detection gas pressures are necessary. Standard operating pressure at present is 30 mbar. In Fig. 3a, the components located inside the gas volume are displayed. The GIC features an anode that is partitioned into four segments. Each segment is 58 mm long, 66 mm wide and separated from neighboring segments by 2 mm gaps. Both cathode and anode are fastened with insulating screws on a frame. A part of the frame that is 11 mm wide, referred to below as the grid holder, covers the edges of the anode segments. The width of the central part of the anode not covered by the grid holder is 44 mm. The length of the non-covered part of the first and last anode segments is 47 mm. As such, the total length of the detection volume is 216 mm. A Frisch grid [22] is mounted 13 mm above the anode, on top of grid holder. From the grid to the cathode, the height of the detection volume is 59 mm. The cathode, all individual parts of the anode and the frame with the Frisch grid are electrically accessible from the outside of the detector via BNC lead-throughs. Fig. 3b shows an entry membrane on the holder used to mount it in the detector. In order to get ions to enter the GIC at the position where the first anode segment starts, the entry membrane is raised 22 mm from the surface of the holder.

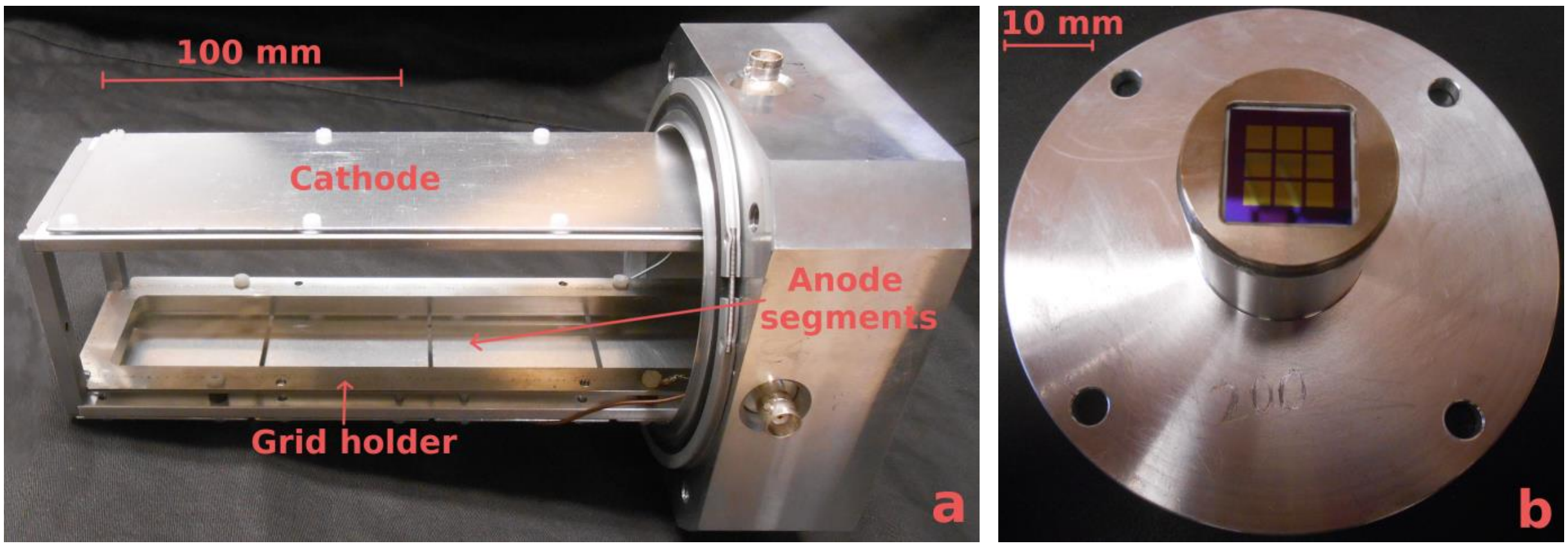

**Figure 3**: Components of the gas ionization chamber.

(a) Cathode, anode and Frisch grid as mounted in the detector. The Frisch grid is located on top of the grid holder, but it is hard to make out in the figure as the individual ribs that make it up are very thin to maximize the open area fraction.

(b) Entry membrane on holder.

Under typical operation conditions, the cathode and the outer shell of the detector are grounded, whereas the Frisch grid and anode are kept at 90 V and 210 V respectively. Output signals are taken between the anode and ground using sets of 142C preamplifiers and 572A analog main amplifiers from Ortec for every anode segment of interest. Several anode segments may be connected in parallel externally and used with a single such amplifier set. The energy resolution of the GIC is on the order of 50 keV and depends on ion species, energy, as well as the number of anode segments used. Energy straggling in the entry window (primarily for heavy ions) and electronic noise affect the energy resolution.

**3. Ion species separation**

Since the purpose of ToF-HIERDA measurements is to obtain elemental depth profiles for the first micrometer below the sample surface, the detector must provide two types of information: (a) identification of ion species and (b) accurate measurement of the ions' energy. Using this information, codes such as Potku [23] can be employed to calculate the sought profiles, taking collision cross sections and stopping power in the sample matter into account. The energy signal from the GIC is not used directly to get a value for the energy of detected ions. Instead, the ToF signal is used to calculate the energy once the mass of an ion is known. Thus, a better energy resolution, and therefore a finer depth resolution in the calculated depth profile, is obtained. Identification of ion species is achieved by placing restrictions on the values of the up to five available signals, i.e. ToF plus the four partial energy signals from the GIC anode segments. Manual selection of regions of interest in two dimensional plots of combinations of signals is often sufficient to separate all species present in a sample. This treatment, and the creation of all figures presented here, has been performed with our own code, employing the matplotlib 2D plotting library for Python. The simplest approach is to use only two signals; time-of-flight and full energy. This is referred to as "full energy mode" below. Full energy is measured by connecting in parallel as many anode segments in the GIC as is necessary to cover the region in which ions of interest are stopped, but not more as electronic noise is minimized by minimizing the total anode length. Separation in this mode is based on ion mass since the full energy of ions is compared to their velocity. Fig. 4 provides an example of the result from a measurement in full energy mode, with a range of data sets from several reference samples that have been superimposed. The reference samples were selected to give an impression of the mass resolution obtained for light elements in the range from Be to Si, as well as to show the separation of a few relevant heavy elements: Mo, Ag, I, W and Au.

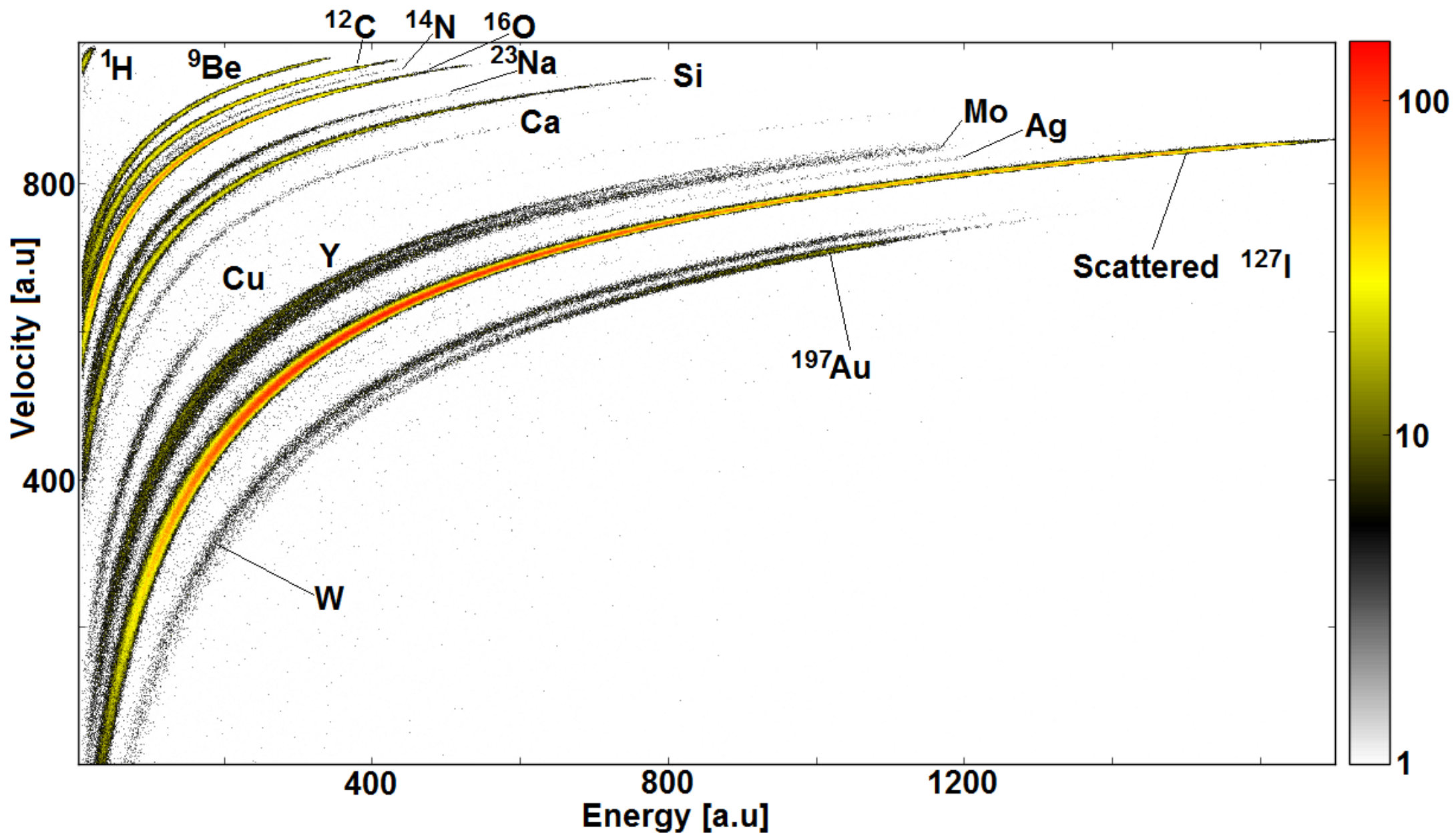

**Figure 4:** Added ToF-E spectra from 36 MeV $^{127}I^{8+}$ ERDA on reference samples of Be, Si, SiO, YO, Mo, W, 22 karat leaf Au alloyed with Ag on $Na_2O$-CaO-$SiO_2$ glass as well as a film containing C, H, and Au on Cu. Energy was recorded with anode segments 1 to 3 in the GIC and single count background has been removed.

A common situation where full energy mode operation is insufficient is when $^{15}N$ is to be quantified in oxygen rich samples. Fig. 5a shows a typical spectrum from a carbon sample, with a number of light isotopes present in the top micrometer layer, obtained using a ToF measurement as described above combined with a silicon solid state detector for the energy signal. The data was taken from the measurements presented in ref. [15]. The displayed part of the spectrum is similar to that obtained with the gas ionization chamber in full energy mode. We are interested here in separating the signals from $^{15}N$ and $^{16}O$. At the tip of the curves, corresponding to the sample surface, the two species are separated. Moving towards lower energies, however, this becomes increasingly difficult. Hence, the sensitivity and depth range of ToF-HIERDA for quantifying small amounts of $^{15}N$ is limited when a large background of $^{16}O$ is present. The maximum depth range with the solid state detector system in this case has been estimated to be approximately 400 nm [15]. To remedy the problem, one can instead attempt separation via the partial energy signals from different anode segments in the GIC ("partial energy mode"). Such treatment of the data provides separation of ions based on atomic number, because it relies on stopping power in the detection gas. Fig. 5b shows a spectrum where $^{9}Be$, $^{12}C$, $^{15}N$ and $^{16}O$ are separated in this way, by plotting the signal from the first anode segment against the one from the remaining three segments connected in parallel. The figure was produced by adding spectra from the scattering of $^{15}N$ and $^{16}O$ beams from a solid W sample and an Au film on glass and performing HIERDA with a 36 MeV $^{127}I^{8+}$ beam on samples of $^{9}Be$ and natural C. In Fig. 5c, the separation of $^{15}N$ and $^{16}O$ has been further optimized by increasing the detector pressure and reducing the anode length, using only segments 1 and 2, while the remaining segments were grounded. In partial energy mode, resolution for neighboring isotopes can be optimized by fine-tuning the detector pressure to match the ion species and energy of interest. If the respective stopping curves for two isotopes have their maxima at distances corresponding to different anode segments in the detector, one can use the ratio of the signals from those anode segments to identify counts. A case-by-case analysis to determine the optimal pressure for a specific pair of isotopes, and selection of active anode segments, is necessary to provide the best possible resolution. Both in Figures 5b and 5c, separation between the displayed ion signals is achieved over a wide energy range. It should be noted, however, that the method can only be applied when the energy of ions is high enough so that they reach at least the second anode segment in the GIC before being fully stopped.

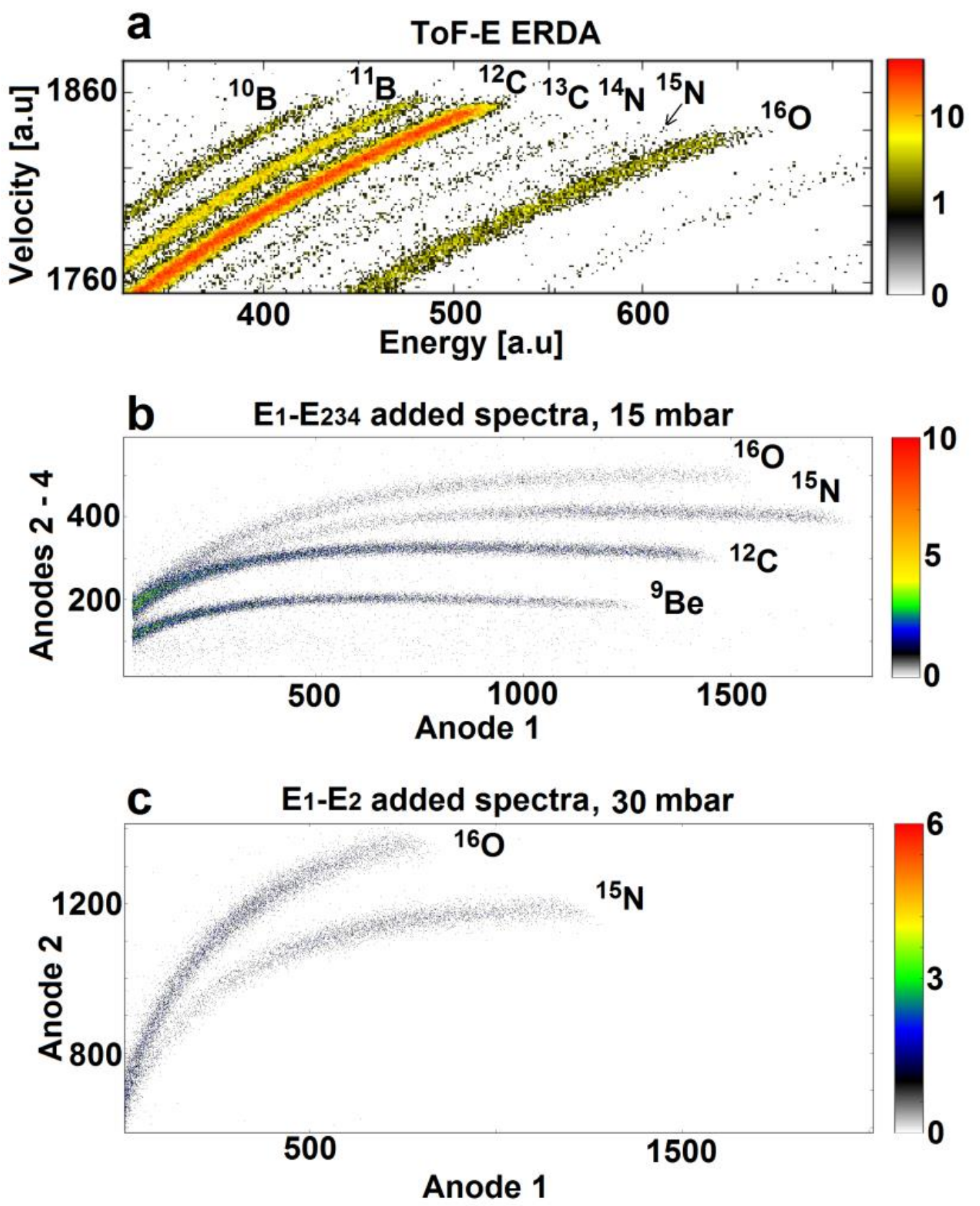

**Figure 5:** Light element separation with ToF-E using a solid state detector compared to using GIC partial energy signals.

(a) ToF-E spectrum from ERDA measurement on a carbon sample with surface enrichment of B, $^{14}$N, $^{15}$N and $^{16}$O using a 36 MeV $^{127}I^{8+}$ beam and a solid state silicon detector for the energy signal.

(b) Added $E_1$-$E_{234}$ spectra from scattering of 7.5 MeV $^{15}$N and $^{16}$O from Au and W as well as ERDA with a 36 MeV $^{127}I^{8+}$ beam on $^{9}$Be and $^{12}$C reference samples. Detector pressure: 15 mbar.

(c) Added $E_1$-$E_2$ spectra from scattering of 7.5 MeV $^{15}$N and $^{16}$O from Au and W. Resolution is improved as compared to (b) thanks to the higher pressure of 30 mbar, which allows the use of only half the detector volume, half the anode length and thereby half the amount of induced electronic noise.

Returning to the need for separating heavy elements, a comparison between the performance of the present detection system and the silicon detector counterpart for that task is provided in Fig. 6. The shown ToF-HIERDA spectra are typical for molybdenum diagnostic mirrors from the First Mirror Test program at JET. Fig. 6b is based on data from ref. [14] using a silicon detector. For Fig. 6a, we have repeated the measurement on a similar sample with the present system in full energy mode. The problem of inadequate separation of Mo and W from forward scattered I originating from the analyzing beam is demonstrated in Fig. 6b. Comparing to Fig. 6a, we see that the use of a GIC for the energy signal completely eliminates that problem. The reason is that a GIC, due to a larger detection volume than a silicon detector, is not as prone to pulse height defects due to charge carrier recombination. As a result, the energy resolution of the GIC for ions that are stopped over a short distance, i.e. heavy ions or ions with low energy, is expected to be better than that of a silicon detector. Pulse height defects are described in detail in ref. [24] and the references therein. Fig. 6a shows that full energy mode operation is adequate for heavy element separation in this case.

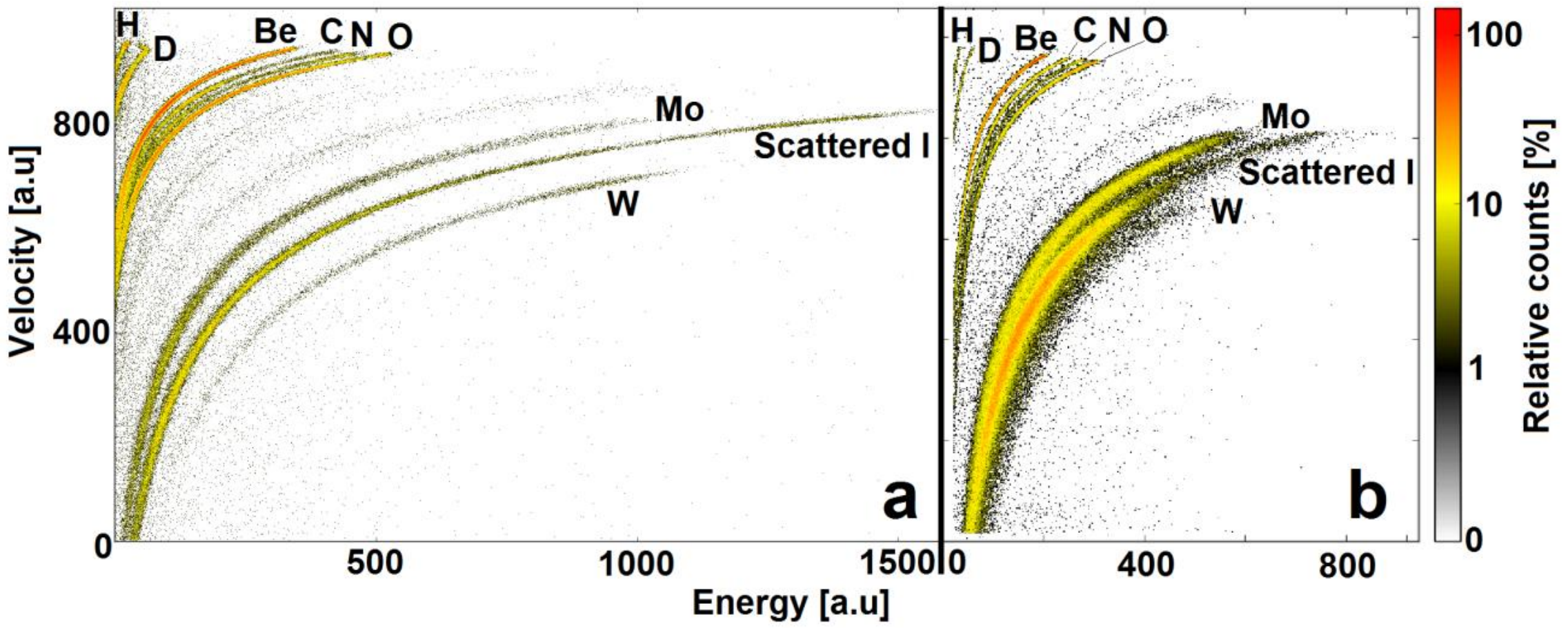

**Figure 6:** 36 MeV $^{127}I^{8+}$ ToF-HIERDA spectra from molybdenum mirrors with H, D, Be, C, N and O present in the top micrometer layer.

(a) Using the new ToF-GIC system in full energy mode.

(b) Using a solid state silicon detector for the energy signal.

## 4. Summary and concluding remarks

A ToF-HIERDA setup has been developed using a GIC for the energy signal. The GIC features a segmented anode that can be used to operate it in partial energy mode for enhanced ion species separation based on atomic number. This has been demonstrated for the case of separating $^{15}N$ from $^{16}O$, which is highly relevant for tracer experiments in fusion research. The methodology can be extended to the separation of other tracer isotopes, for example $^{18}O$ from $^{19}F$ or $^{22}Ne$ on ceramic components containing $^{23}Na$. Other applications include studies of hydrogen and deuterium diffusion, nitrogen quantification ($^{14}N$ and $^{15}N$) for metallurgy or surface processing and the study of surface modification in samples by ion implantation. When studying light elements, in particular H and He, the detection efficiency of the time-of flight-system must be considered in order to correctly calculate elemental concentrations from raw time-of-flight/energy spectra.

Since the selection of anode segments and detector pressure can be independently varied, ions of a specific mass, nuclear charge and energy can be forced to deposit the majority of their energy over a selected anode segment. As such, it is in principle possible to tune the detection system in partial energy mode for separating a selected pair of ion species optimally. The fact that the system delivers sufficient mass resolution to separate light elements in the range from H to Ne and heavy elements such as Mo, I and W simultaneously in full energy mode makes it ideal for the study of metal mirrors, as the correct identification of heavy elements is needed for absolute quantification of deposited light species without dependence on a beam current measurement.

The system was shown to provide better separation of signals from different species than a similar system using a solid state silicon detector. This is true for heavy elements when operating in full energy mode (separation based on ion mass) and for light elements in partial energy mode (atomic number). In addition, a gas ionization chamber does not suffer from radiation damage, whereas as a solid state detector does. One could continuously replace the gas if resolution deterioration due to breakdown of molecules by the effect of incoming ions poses a problem. The spectra that have been presented here, however, were all obtained without replacing the gas over several hours of operation.

**Acknowledgments**

We would like to acknowledge the contributions of Jonas Åström and Rogério Zorro at the Tandem Laboratory, Uppsala University and Håkan Ferm at KTH who were taking part in, and guiding us through, many practical tasks. These included assembly of the mechanical structure of the detector (vacuum components, support beams and stabilization), leak testing, selecting and connecting electronics and constructing the gas supply system. Furthermore, we want to mention Daniel Primetzhofer with whom we had many fruitful discussions and Valentina Paneta for taking part in the mounting of microchannel plates in the time-of-flight detectors as well as verifying the composition of the reference sample of C, H, Cu and Au via PIXE measurement. Alvaro Garcia-Carrasco at KTH took part in the measurements behind Fig. 6a and supplied the corresponding data. We are thankful to Mike Watkins and Armin Weckmann for proofreading the manuscript.

This work has been carried out within the framework of the EUROfusion Consortium and has received funding from the European Union's Horizon 2020 research and innovation programme under grant agreement number 633053. The views and opinions expressed herein do not necessarily reflect those of the European Commission.